\documentclass[12pt,preprint]{aastex} 

\def \etal{{\rm et al.}} 
\def \eg{{\rm e.g.,}} 
\def \ie{{\rm i.e.,}} 
\def \TRACE{{\it TRACE}} 
\def \SOHO{{\it SOHO}}

\shorttitle{CDS Coronal Loops} 
\shortauthors{Schmelz and Martens}

\begin{document}

\title{ Multithermal Analysis of a CDS Coronal Loop}

\author{J.T. Schmelz } 
\affil{Physics Department, University of Memphis, Memphis, TN 38152} 
\email{jschmelz@memphis.edu } 
\and 
\author{P.C.H. Martens}
\affil{Physics Department, Montana State University, Bozeman, MT 59717}
\begin{abstract}

The observations from 1998 April 20 taken with the Coronal Diagnostics Spectrometer (CDS) on the {\it Solar and Heliospheric Observatory} (\SOHO) of a coronal loop on the limb have shown that the plasma was multi-thermal along each line of sight investigated, both before and after background subtraction. The latter result relied on Emission Measure (EM) Loci plots, but in this Letter, we used a forward folding technique to produce Differential Emission Measure (DEM) curves. We also calculate DEM-weighted temperatures for the chosen pixels and find a gradient in temperature along the loop as a function of height that is not compatible with the flat profiles reported by numerous authors for loops observed with the EUV Imaging Telescope (EIT) on \SOHO\ and the {\it Transition Region and Coronal Explorer} (\TRACE). We also find discrepancies in excess of the mathematical expectation between some of the observed and predicted CDS line intensities. We demonstrate that these differences result from well-known limitations in our knowledge of the atomic data and are to be expected. We further show that the precision of the DEM is limited by the intrinsic width of the ion emissivity functions that are used to calculate the DEM, which for the EUV lines considered is of the order dlog(T) ~ 0.2 - 0.3.  Hence we conclude that "peaks and valleys" in the DEM, while in principle not impossible, cannot be confirmed from the data.

\end{abstract}


\section{Analysis}

Solar active regions are composed of a variety of loops that are revealed when closed magnetic structures are filled with high-temperature, low-density plasma. The actual heating mechanism of this plasma remains a source of ongoing investigation, but the temperature structure of these loops is thought to provide clues to its elusive nature.

The \SOHO -CDS instrument (Harrison \etal\ 1995) observed a loop on the southwest limb on 1998 April 20 at 20:54 UT. The spectral data and the original analysis are described in detail by Schmelz \etal\ (2001) and the background subtraction was done by Schmelz \etal\ (2005). Figure 1 shows multiple wavelength frames for this loop, which is seen most clearly in the hotter temperature lines. There is certainly cool plasma above the limb, and the analysis described below will determine if it is part of the main loop structure. The loop alignments from various spectral lines were well within the instrument point spread function (see below). We chose three pixels along the visual center of the loop for detailed analysis, one near the apex, one at the upper part of the southern loop leg, and a third in the lower portion of the southern loop leg. We also selected a pair of background pixels for each loop pixel, one inside the loop and one outside. These positions are the same as those shown in Figure 2 of the Schmelz et al. (2005) paper.

We use the calculations compiled in version 4.02 of the CHIANTI Atomic Physics database (Dere \etal\ 1997; Young \etal\ 2003), the ionization fractions of Mazzotta \etal\ (1997), and the ``hybrid'' elemental abundances of Fludra \& Schmelz (1999). The spectral lines available in this data set are listed in Table 1, where the columns show a running line number, the ion, its wavelength in \AA, the log of its peak formation temperature, and the spectral line intensities and uncertainties for the three loop pixels, both before and after background subtraction. Note: CHIANTI 5.1 makes a few small changes to three of our lines: O~V $\lambda$629.73 ($<$ 5-10\% at maximum), and Fe XII $\lambda$346.85, $\lambda$364.46 ($<$ 30\% at maximum), but these will have no impact on our results (Landi 2005, private communication).

In coronal equilibrium, the intensity $I$ of an optically thin spectral line of wavelength $\lambda$ is given by:

\begin{equation}
I(\lambda_{ij}) = {1 \over 4 \pi}\ {hc \over \lambda_{ij}}\ A \times\   
\int_0^\infty G(T)\,DEM(T) dT.
\end{equation}

\noindent
where the units of $I$ are ergs\,cm$^{-2}$\,s$^{-1}$\,sr$^{-1}$, 
$h$ is Planck's constant, $c$ is the speed of light, $A$ is the elemental abundance,
$G(T)$ is the contribution function, and $T$ is the electron temperature.
The DEM is defined via$\int G(T)\,{n_e^2}\,dl = \int G(T)\, DEM(T) dT$, where
$n_e$ is the electron density and $l$ is the line-of-sight element of length in cm.

Rather than using inversion to determine the DEM, we use a forward-folding
approach in the present analysis. We begin with a flat source function 
of log DEM = 21.5\,cm$^{-5}$. This initial model is folded through the 
spectral line emissivities to produce a set of predicted intensities that are compared 
with the observed values. The DEM is then adjusted manually in small steps 
to improve the agreement between the observed and 
predicted intensities while the curve is kept as smooth a function of 
temperature as possible. The process is repeated until, ideally, the 
predicted and observed intensities agree to within approximately $\pm$1-2$\,\sigma$ 
of the observed values. We also make use of density-sensitive lines by running the 
DEM program with different density values in the range of 1e8, 2e8, 3e8, . . . , 1e10 
cm$^{-3}$. In each independent run, the DEM curve can be adjusted (if need be) to 
improve the fit to the data. The best-fit densities resulting from this method were 1e9, 3e9, and 5e9 cm$^{-3}$ for the apex, upper-leg, and lower-leg pixels, respectively. 

Figure 2 shows the resulting DEM curves both before (dashed) and after (solid)
background subtraction. The DEM-weighted temperatures, 
$ T_{DEM}\ =\ \sum_{i}\ (T_i\ \times\ DEM(T_i))\/ \ \sum_{i}\ (DEM(T_i))$, 
are plotted as a function of the arc length along the loop in Figure 3. These
background-subtracted results show even greater curvature than the original
results, so these data are now even less in agreement with the flat temperature 
profiles reported by numerous authors for loops observed with EIT 
(Neupert \etal\ 1998) and \TRACE\ (Lenz \etal\ 1999). 

One important issue related to this work involves the pixel sizes and point spread 
functions (PSF) of \TRACE\ and CDS. \TRACE\ has $0.52''\times 0.52''$ pixels and a 
matching PSF FWHM of about $1''$. CDS has $4''\times 1.6''$  pixels with the 4$''$ slit. 
Harrison et al. (1995) measured a CDS pre-launch PSF of $1.2''\times 1.5''$. Pauluhn 
et al. (1999) measured the PSF in-orbit; if they assume a functional form similar to 
the one measured by Harrison et al. (1995), they obtain $3''\times 4''$. Some CDS 
pixels will fall entirely within an average \TRACE\ loop, which has a typical cross 
section of 5 to 10 pixels (Watko \& Klimchuk 2000), so for either value of the CDS 
FWHM the background-subtracted spectrum from these pixels would indeed be a true 
\TRACE\ loop spectrum (\eg\ Cirtain 2005). Therefore, it is not necessarily true 
that the flat-temperature profiles determined from EIT and \TRACE\ image ratios 
can be explained by the smaller pixel size. We feel that it is more likely that 
these \TRACE\ temperature profiles result from multithermal plasma along the line 
of sight (Martens \etal\ 2002; Schmelz 2002), consistent with the CDS DEM curves 
presented here.

\section{Discussion}

Aschwanden (2002) asserted that the relatively flat DEM distributions
constructed by Schmelz \etal\ (2001) for the non-background subtracted CDS data
were an artifact of an over-zealous temperature smoothing function. Since this
paper is a follow-up to the Schmelz \etal\ paper and uses the same analysis
techniques, it is important to clarify this confusion.

Aschwanden (2002) used the same pre-background subtracted CDS data analyzed by
Schmelz \etal\ (2001) for the 1998 April 20 loop. His resulting DEM curves
looked somewhat different, however, because he did not apply a smoothing
function. Rather, he simply allowed the data to dictate the shape of the DEM,
resulting in a distribution with several peaks and valleys. Furthermore, he claimed that
the resulting peaks and valleys revealed the actual temperatures of various
isothermal loops along the line of sight, and that these temperature values
could be recovered by the standard \TRACE\ narrow passband ratio analysis. This
last point is {\it not} correct, as the following example illustrates. Suppose a DEM
distribution showed peaks at 1.0 and 1.3 MK and that these peaks represented
two separate loops along the line of sight. The signal from the \TRACE\ 171-\AA\ 
passband will be dominated by the contribution from the first loop and that of 
195-\AA\ passband by the second loop. The ratio will then yield a single 
temperature somewhere between 1.0 and 1.3 MK, which does not represent the 
actual temperature of either loop. Therefore the single filter ratio method used 
to determine loop temperatures is inadequate with either a smooth DEM distribution 
or a distribution with multiple peaks.

Next we discuss the issues related to a smooth DEM curve versus 
a curve with multiple peaks and valleys. 
In the original DEM curves of Schmelz \etal\ (2001) and in the
background-subtracted curves presented here, there are discrepancies between the
observed and predicted line intensities of several sigma for a few lines. This
is more than one would expect for a good mathematical fit to the data. It is
also true that a peaks-and-valleys DEM distribution would result in better
fits and lower values of reduced $\chi^2$. It is important to note, however,
that deviations of several sigma are {\it expected} for 
some spectral lines in DEM analysis, not only because of blends but also 
because the atomic data (although it continues to improve) still has limitations.

It is widely known that many coronal EUV emission lines do not behave as
predicted, even under circumstances where the line-of-sight temperature is well
defined (\ie\ the DEM distribution is narrowly peaked). This is especially true
for lines from the lithium and sodium isoelectronic sequences (Landi, Feldman \& Dere 2002a,b; Del Zanna, Landini \& Mason 2002). We have four Li-and Na-like lines in our data set. The Ca X line at 557 \AA\ is part of the loop background, so it does not affect the background-subtracted DEM curves presented here. The Mg X line at 624 \AA, however, is visible in the upper and lower leg pixels after subtraction, but the fits are so bad that we choose not to use it to produce the DEM curves. We get better agreement for Si XII at 520 \AA\ (Li-like) and Fe XVI at 360 \AA\ (Na-like), however, which are well fit by these DEM curves. They were constrained on the high temperature end by the SXT data in the original analysis (Schmelz et al. 2001), and we see no reason to reject them based on the data themselves. Several experts we consulted thought that the problem might apply to all Li- and Na-like lines, and others thought that the problems might be more serious for transition region lines than for coronal lines. Our data seem consistent with the latter view, but it is clear that more work is needed on this important issue.

Numerous authors have observed iron lines that do not conform to predictions (see, e.g., Brosius et al. 1996; Young, Landi \& Thomas 1998; Binello \etal\ 2001; Landi et al. 2002a,b;). Many of these iron lines (including some available in this data set) are not recommended for DEM analysis because (1) they are blended with other lines, many unidentified; (2) they have a strong density dependence; or (3) there are unresolved issues with theoretical emissivities. We have been able to get around some of these problems by avoiding lines with known blends and making use of density sensitive lines with our forward-folding method. We also note that new atomic data have been included in upgrades of CHIANTI, and the agreement between the observed and predicted intensities for some of the lines used here has improved since we did the original analysis with version 2.0.

Fludra \& Sylwester (1986) showed that even in an idealized, error-free
situation, if one were to over-sample the temperature axis, some DEM methods 
will produce an oscillating solution, \ie\ one with peaks and valleys. Lanzafame 
\etal\ (2002) criticized DEM inversion techniques and discussed the reasons for 
spurious peaks and valleys in the resulting temperature distributions. The main 
problem is still potential errors in the atomic data. These errors are difficult to 
quantify, and are rarely if ever included in the inversion. Any measurement error 
in the instrumental calibration, elemental abundances, or ionization balance could 
also create spurious DEM features. A. Fludra (2003, private communication) 
points out that one need look no further than the progression in recent
years of the ionization balance calculations. It seems only likely that these values 
will change again.

Aschwanden (2002) derived DEM curves directly from the density and temperature
distribution in a sophisticated model atmosphere simulation, rather than from
the spectroscopic line intensities that such a model would generate. Hence
his DEMs have in principle unlimited temperature resolution. Aschwanden then
downgraded the DEM resolution by binning in intervals of dlog(T) = 0.1. In
reality, however, the resolution of DEM curves is naturally limited by the
prevalent widths of the ion emissivity functions, which are of the order
dlog(T) = 0.2 - 0.3 (\eg\ Mazzotta \etal\ 1998). This is analogous
to the problem of the resolution of a telescope, which is determined by the
width of the instrument's point spread function (as long as the pixel size
satisfies the Nyquist criterion). In fact, the mathematical
formulation of the two problems is nearly identical. It is well known that
any feature in a telescope image with a size smaller than the width of the
point spread function must be regarded with the utmost skepticism, and the
same is true for any feature in a DEM with dlog(T) $<$ 0.2 - 0.3. Hence
even if Aschwanden's (2002) model DEMs were correct -- and they must be,
since they are derived directly from the model density and temperature
distribution -- they could not be recovered using real spectral data.

We avoid many of these problems by using forward folding rather than inversion, 
but any DEM result is only as good as the atomic data that go into it.
DEM peaks and valleys are often artifacts of the mathematical inversion 
procedure or of underestimating uncertainties in DEM analysis. Since atomic 
physics errors are known to cause spurious features, such structures should be 
treated with caution and should be interpreted as real only when there is redundant 
evidence that they exist.

\section{Conclusions}

In this Letter, we confirm the results of Schmelz \etal\ (2005) who found that the background-subtracted intensities for the 1998 April 20 CDS loop were not consistent with isothermal plasma along the lines of sight investigated. With their EM Loci plots, however, they could not tell how the temperature varied along the loop length. Here we have determined full DEM distributions for the plasma at the same three loop positions. These results show that the DEM-weighted temperature profile is compatible with a steep transition region and slow temperature rise in the corona, and not compatible with the flat profiles reported by numerous authors for loops observed with EIT and \TRACE. 

Do these EIT and  \TRACE\ loops have sharply peaked temperature distributions or broadly peaked DEM plateaus? In either case, the passband ratios produce incorrect plasma temperatures (Martens \etal\ 2002; Schmelz \etal\ 2003). We also conclude that the deviations of several sigma between the observed and predicted line intensities are expected because of the uncertainties in the atomic data, and the finding that \TRACE\ loops are isothermal along their axis should be regarded with a strong dose of skepticism. 

\acknowledgements

We would like to thank K. Nasraoui and J. Cirtain. Solar physics research at the University of Memphis is supported by NSF ATM-0402729 and NASA NNG05GE68G. The data analyzed here are courtesy of the \SOHO\ CDS consortium. \SOHO\ is a project of international cooperation between ESA and NASA. CHIANTI is a collaborative project involving the NRL (USA), RAL (UK), and the Universities of Florence (Italy) and Cambridge (UK). \TRACE\ mission operations and data analysis at Montana State University are supported by NASA through a subcontract with Lockheed-Martin.

{}

\clearpage

\begin{figure}
\figurenum{1}
\epsscale{.9}
\plotone{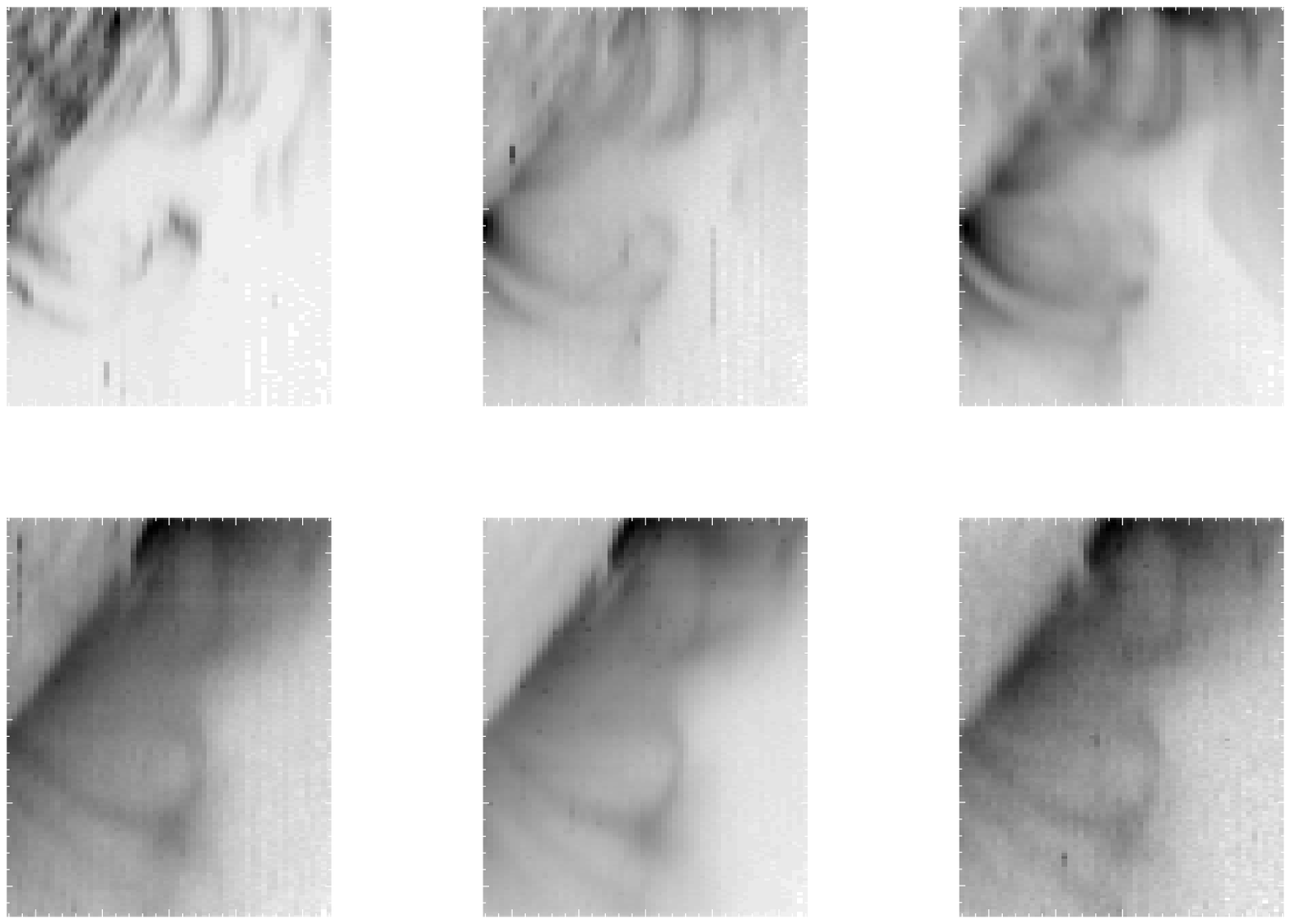}
\caption{Co-aligned CDS loop data from 1998 April 20 at 20:54 UT seen here on the southwest limb. The images are arranged in order of peak formation temperature, from coolest to hottest (see Table 1 for details): O V (629 \AA), Ca X (557 \AA), Mg IX  (368 \AA), Al XI (568 \AA), Si XII (520 \AA), and Fe XVI (360 \AA). All frames depict the monochromatic peak intensity of the ion, not the summed intensity over the entire CDS wavelength window. The intensity scale has been inverted so the loops appear dark in these images. Cosmic ray hits were flagged as missing data and were not included in the analysis.}
\end{figure}

\clearpage

\begin{figure}
\figurenum{2}
\epsscale{.9}
\plotone{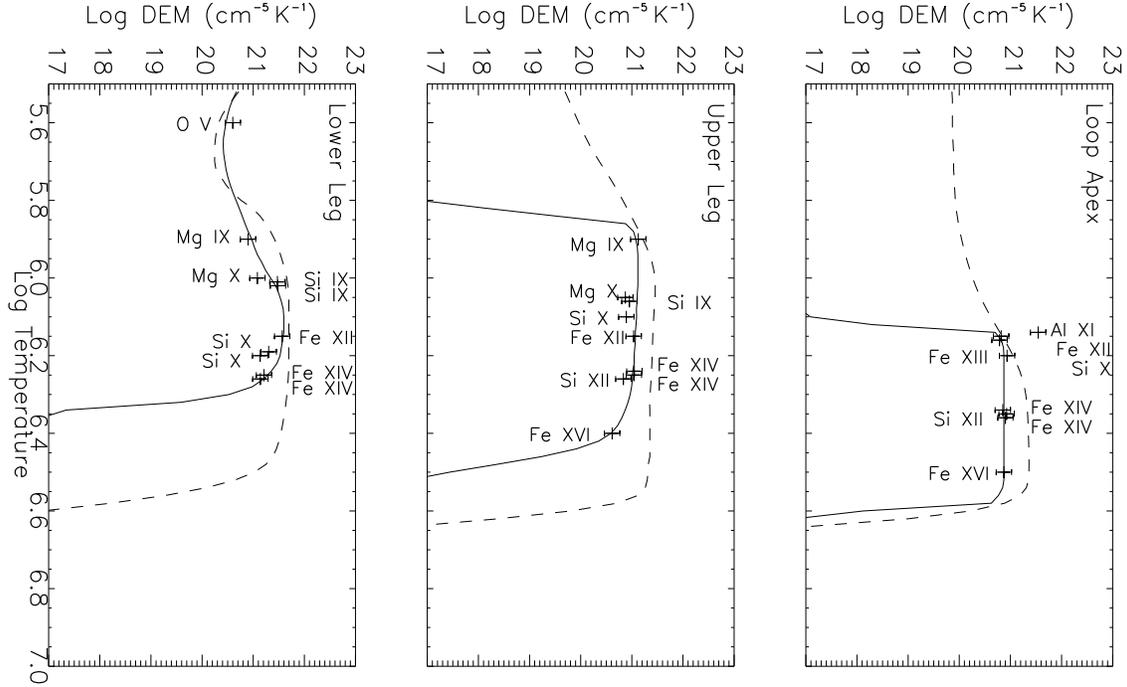}
\caption{ The DEM curves for the three chosen pixels before (dashed) and after (solid) 
background subtraction. All curves are constrained on the high-T end by co-spatial, 
co-temporal SXT data; the top two curves are constrained on the low-T end by the 
subtracted CDS data. The CDS points are plotted at the position of their maximum 
contribution to the DEM. The error bars are the fit uncertainties from Table 1. 
Two discrepant points were not used to create the curve: the Mg~X line at 624 \AA\ 
(line 6) is a Li-like line known to give problems (see text) and the most probable 
cause for the bad fit to the Al~XI line at 568 \AA\ (line 11) is contamination from 
the strong, second-order Fe XV line at 284 \AA. Note the good agreement for Li-like 
Si XII at 520 \AA\ (line 20) and Na-like Fe XVI at 360 \AA\ (line 21).}
\end{figure}

\clearpage

\begin{figure}
\figurenum{3}
\plotone{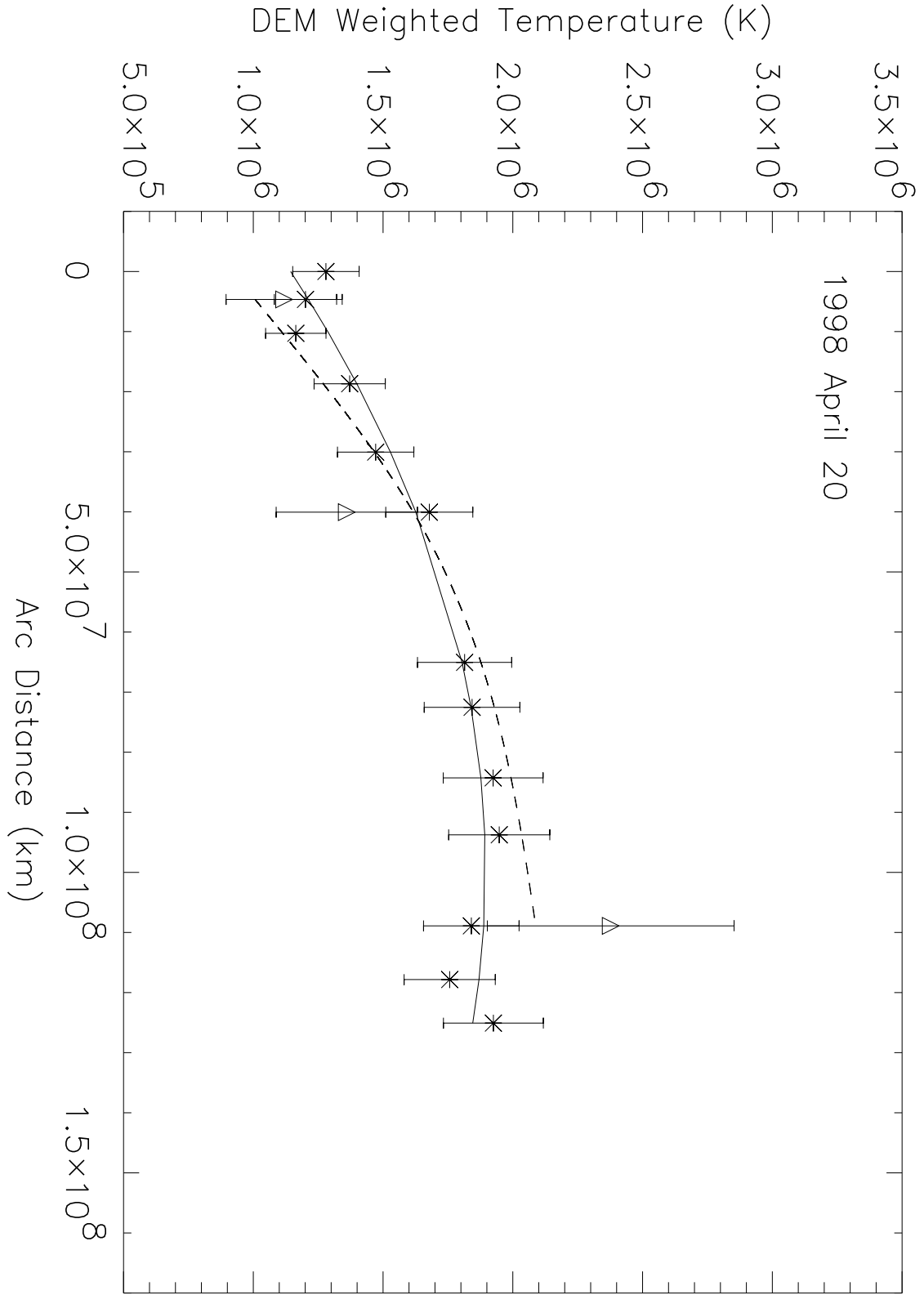}
\caption{Plot of DEM-weighted temperature vs. arc distance for the 1998 April
20 loop. The stars and triangles represent the pre- and post-background subtracted points, respectively.} 
\end{figure}

\clearpage

\begin{deluxetable}{llllrrrrrr}
\tabletypesize{\scriptsize}
\tablewidth{0pt}
\tablecaption{ Spectral Line Intensities and Uncertainties (ergs cm$^{-2}$ s$^{-1}$ sr$^{-1}$)}
\tablehead{
\colhead{} & \colhead{Ion} & \colhead{$\lambda$ (\AA)} & \colhead{Log T} & 
\colhead{Apex} & \colhead{Subtracted} &
\colhead{Upper Leg} & \colhead{Subtracted} & 
\colhead{Lower Leg} & \colhead{Subtracted}   
}
\startdata

1&O V &629.73 &5.40 &5.18$\pm$0.92&-5.95$\pm$2.24 &13.6$\pm$1.89&5.33$\pm$2.99&322.$\pm$6.63&258.$\pm$8.17\\
2&Ne VI&562.80&5.65&5.11$\pm$1.19&0.16$\pm$2.24 & 14.2$\pm$2.80 & 12.2$\pm$3.16& 32.6$\pm$2.98 & 19.0$\pm$4.09\\
3&Ne VII&561.73&5.70&1.02$\pm$0.73&-1.53$\pm$1.85 & 4.62$\pm$1.81 &4.62$\pm$1.81 & 4.92$\pm$1.82& -2.65$\pm$3.15\\
4&Ca X&557.77&5.80&10.2$\pm$1.42&3.92$\pm$2.15 & 47.9$\pm$2.69&18.5$\pm$4.21&93.1$\pm$3.99&28.8$\pm$6.22\\
5&Mg IX&368.07&6.00&207.$\pm$4.73& 31.1$\pm$7.72 &1022.$\pm$9.76&327.$\pm$15.0&1883.$\pm$13.4&606.$\pm$20.6\\
6&Mg X&624.94&6.05&95.1$\pm$3.60& 28.4$\pm$5.73 &259.$\pm$6.00&65.9$\pm$9.54&428.$\pm$7.80&86.3$\pm$12.6\\
7&Si IX&341.95&6.05&28.2$\pm$2.22&-0.33$\pm$3.96&65.2$\pm$3.55&6.44$\pm$5.78&118.$\pm$5.13&45.1$\pm$7.93\\
8&Si IX&349.87&6.05&20.1$\pm$2.06&-17.1$\pm$4.01 &198.$\pm$5.17 & 57.6$\pm$8.09 & 389.$\pm$7.33 & 161.$\pm$10.9\\
9&Si X&356.01&6.10&46.3$\pm$2.67&14.4$\pm$4.06 &221.$\pm$5.34 & 86.7$\pm$8.18 & 489.$\pm$7.64 & 192.$\pm$11.5\\
10&Al XI&550.03&6.15&25.8$\pm$2.13&12.1$\pm$3.05 &45.1$\pm$2.71 & 8.37$\pm$4.46 & 76.2$\pm$3.70 & 9.87$\pm$6.02\\
11&Al XI&568.12&6.15&56.1$\pm$2.88&29.2$\pm$4.21 &65.6$\pm$3.30 & 13.8$\pm$5.32 & 101.$\pm$4.14 & 20.8$\pm$6.69\\
12&Fe XII&346.85&6.15&24.2$\pm$2.03&9.55$\pm$3.65 &67.9$\pm$5.10 & 18.5$\pm$8.19&102.$\pm$6.92 & 18.5$\pm$11.1\\
13&Fe XII&364.46&6.15&65.2$\pm$3.06&30.5$\pm$4.69 &252.$\pm$5.28 & 69.1$\pm$8.56 &478.$\pm$7.44&181.$\pm$11.6\\
14&Si X&347.40&6.10&77.6$\pm$3.17&37.7$\pm$5.02 &210.$\pm$5.78 & 41.7$\pm$9.19 & 382.$\pm$7.68 & 93.7$\pm$12.4\\
15&Fe XIII&320.81&6.20&35.3$\pm$2.66&19.7$\pm$6.44 &89.5$\pm$7.56&32.4$\pm$10.6&148.$\pm$9.39&71.0$\pm$14.3\\
16&Fe XIII&321.40&6.20&21.6$\pm$2.84&8.49$\pm$4.60 &27.8$\pm$4.56&7.27$\pm$6.95&18.5$\pm$6.10&-10.8$\pm$9.69\\
17&Fe XIII&348.18&6.20&78.7$\pm$3.31&44.0$\pm$4.95 &114.$\pm$5.20&40.8$\pm$8.28&176.$\pm$7.06&48.6$\pm$11.4\\
18&Fe XIV&334.17&6.25&181.$\pm$5.12&83.8$\pm$7.57 &328.$\pm$6.66
&100.$\pm$10.4&571.$\pm$8.40&132.$\pm$13.7\\
19&Fe XIV&353.83&6.25&60.4$\pm$3.13&27.5$\pm$4.42 &131.$\pm$4.43&44.7$\pm$6.80&222.$\pm$5.67&67.9$\pm$8.93\\
20&Si XII&520.67&6.25&192.$\pm$5.18&88.7$\pm$7.47 &225.$\pm$5.54&54.0$\pm$8.81&342.$\pm$7.00&38.6$\pm$11.6\\
21&Fe XVI &360.76&6.40&906.$\pm$9.35&296.$\pm$14.4 &922.$\pm$9.43&143.$\pm$15.4&1080.$\pm$10.6&-46.3$\pm$18.2\\

\enddata

\end{deluxetable}

\end{document}